\documentclass[12pt]{article}
\usepackage{amsmath}
\usepackage{graphicx}
\usepackage{enumerate}
\usepackage[round]{natbib}
\usepackage{setspace} 
\usepackage{algorithm}
\usepackage{algorithmic}
\usepackage{xcolor}
\usepackage{url} \usepackage{amssymb}  
\usepackage{mathtools}

\newcommand{\blind}{1}

\begin{document}

\def\spacingset#1{\renewcommand{\baselinestretch}%
{#1}\small\normalsize} \spacingset{1}

%%%%%%%%%%%%%%%%%%%%%%%%%%%%%%%%%%%%%%%%%%%%%%%%%%%%%%%%%%%%%%%%%%%%%%%%%%%%%%\\
%\author{Emma Prevot \hspace{.2cm}\\
%    Department of Statistics, University of Oxford\\
%    and \\ \\
%    Dieter Häring \\
%    Novartis Pharma AG, Basel, Switzerland \\
%    and \\ \\
%    Thomas E. Nichols \\
%    Big Data Institute, Li Ka Shing Centre for Health Information and Discovery \\
%    Centre for Integrative Neuroimaging (OxCIN), FMRIB, \\
%    Nuffield Department of Clinical Neurosciences \\
%    and \\ \\
%    Chris C. Holmes \\
%    Department of Statistics, University of Oxford \\
%    Big Data Institute, Li Ka Shing Centre for Health Information and Discovery \\
%    and \\ \\
%    Habib Ganjgahi \\
%    Department of Statistics, University of Oxford \\
%    Big Data Institute, Li Ka Shing Centre for Health Information and Discovery \\ \\}
%  \maketitle
%}

\newcommand\iidsim{\stackrel{\mathclap{iid}}{\sim}}
\newcommand\iidgibbs{\stackrel{\mathclap{Gibbs}}{\sim}}
\date{}

\if1\blind
{
  \title{\bf A hierarchical modelling approach for Bayesian Causal Forest on longitudinal data: A Case Study in Multiple Sclerosis
}

\author{Emma Prevot $^{1,2 \ast}$, Dieter A. Häring $^{3}$, Thomas E. Nichols $^{2,4}$, \\ Chris C. Holmes $^{1,2}$,  Habib Ganjgahi $^{1,2 }$\\
{\small $^{1}$ Department of Statistics, University of Oxford, 24-29 St Giles', Oxford OX13LB, UK} \\
{\small $^{2}$ Big Data Institute, Li Ka Shing Centre for Health Information and Discovery,}\\
{\small Nuffield Department of Medicine, University of Oxford, Old Road Campus, Oxford OX3 7LF, UK}\\
{\small $^{3}$ Novartis Pharma AG, Basel, Switzerland}\\
{\small $^{4}$ Centre for Integrative Neuroimaging (OxCIN), FMRIB, Nuffield Department} \\ 
{\small of Clinical Neurosciences, University of Oxford, Oxford OX3 9DU, UK}\\
{\small $^\ast$ Correspondence:  emma.prevot@exeter.ox.ac.uk}
}

%\author{Emma Prevot \hspace{.2cm}\\
%Department of Statistics, University of Oxford\\ \\
%Supervised by:\\
%Prof Chris C. Holmes, Prof Thomas E. Nichols, and Dr Habib Ganjgahi \\ \\} 
    \maketitle

    }
\fi

\if0\blind
{
  \bigskip
  \bigskip
  \bigskip
  \begin{center}
    {\LARGE\bf A hierarchical modelling approach for Bayesian Causal Forest on longitudinal data: A Case Study in Multiple Sclerosis Clinical Trials}
\end{center}
  \medskip
} \fi

\begin{abstract}
Long-running clinical trials offer a unique opportunity to study disease progression and treatment response over time, enabling nuanced questions about how and when interventions alter patient trajectories. However, drawing causal conclusions in this setting is challenging due to irregular follow-up, individual-level heterogeneity, and time-varying confounding.  Bayesian Additive Regression Trees (BART) and their extension, Bayesian Causal Forests (BCF), have proven powerful for flexible causal inference in observational data, especially when treatment effects are heterogeneous and outcome surfaces are non-linear. Yet, both models assume independence across observations and are fundamentally limited in their ability to model within-individual correlation over time. This limits their use in real-world longitudinal settings where repeated measures are the norm. Motivated by the NO.MS dataset, the largest and most comprehensive clinical trial dataset in Multiple Sclerosis (MS), with more than 35,000 patients and up to 15 years follow-up, we develop BCFLong, a hierarchical model that preserves BART’s strengths while extending it for longitudinal analysis. Inspired by BCF, we decompose the mean into prognostic and treatment effects, modelling the former on Image Quality Metrics (IQMs) to account for scanner effects, and introduce individual-specific random effects, including random intercepts and time-dependent slope, with a sparsity-inducing horseshoe prior.
Simulations confirm BCFLong’s superior performance and robustness to sparsity, significantly improving outcome and treatment effect estimation over vanilla BCF. On NO.MS, BCFLong captures clinically meaningful longitudinal patterns in brain volume change, which would have otherwise remained undetected. These findings highlight the importance of adaptively accounting for within-individual correlations and position BCFLong as a flexible framework for causal inference in longitudinal data.

\end{abstract}

%\noindent%
%{\it Keywords:}  Bayesian Additive Regression Trees (BART), Causal Inference, Sparsity Inducing Priors
\vfill

\newpage
\spacingset{1.9} % DON'T change the spacing!
\section{Introduction}
\label{sec:intro}

Multiple Sclerosis (MS) affects approximately 2.9 million people around the world. It is a chronic immune-mediated disease of the central nervous system (CNS) characterized by inflammation, demyelination, destruction of neurons and axons that wire the brain, which ultimately leads to severe physical and cognitive disability \citep{who2021ms, understandingms, dobson2019multiple}. Despite remarkable advances in MS drug development which led to more than 20 approved treatments, there is no cure for MS. The available therapies successfully reduce focal inflammation and recurrent acute exacerbations of neurological dysfunction known as relapse, improve quality of life, and slow disease progression. Understanding and halting the insidious progression that occurs apparently independent of new focal inflammation remains a high unmet medical need. 

Magnetic Resonance Imaging (MRI) offers a minimally invasive method to investigate the pathobiology of MS, offering insights into both focal inflammation and diffuse neurodegeneration. Imaging-Derived Phenotypes (IDPs), such as brain volume loss, quantify the cumulative impact of these processes on the central nervous system and are effective for studying the (subclinical) progression and effects of treatment on the entire brain or on specific brain regions. Accelerated brain tissue loss and disease-related reductions in brain volume have been identified as risk factors for increased disability accumulation \citep{sormani2017assessing, miller2018brain, taschler2024normative}. These IDPs are increasingly used in clinical trials as secondary or exploratory endpoints to assess disease activity and treatment efficacy. Compared to conventional clinical assessments, they offer a more objective quantification of disease progression and hold promise for stratifying patients and tailoring therapeutic strategies. In this context, accurately modelling treatment effects versus placebo and identifying brain-related biomarkers that drive heterogeneity in treatment response are critical steps toward advancing personalized medicine in MS.

\subsection{Estimating Causal Effects from Longitudinal Data}

Estimating causal effects is a central challenge in many scientific disciplines, including medicine, economics, and the social sciences. In randomized controlled trials (RCTs), causal effects can be estimated directly because treatment assignment is randomized, thereby balancing both observed and unobserved confounders. However, in observational studies, treatment assignment is often influenced by confounding variables, making causal inference more complex \citep{rubin1978bayesian, pearl2000models}. To formalize causal questions and define causal effects, the potential outcomes framework \citep{rosenbaum1983central, austin2011introduction} is widely used. In this framework, each individual has two potential outcomes, one corresponding to receiving treatment and the other to not receiving it, but only one is observed. The fundamental challenge lies in estimating the missing, unobserved, counterfactual outcome to understand causal treatment effects in observational data. This requires strong assumptions and appropriate methods to approximate what would have happened under the alternative scenario.

Longitudinal causal inference introduces further complexities beyond the cross-sectional setting, including time-varying confounding, where past treatments may influence future confounders and outcomes; dynamic treatment regimes, in which treatment assignment can vary across time and depend on intermediate variables; and intricate within-subject dependencies, where repeated observations on the same individual induce correlation both within and across time points \citep{robins1997causal, hernan2010causal}. These challenges complicate both identification and estimation. Established methods include inverse probability weighting (IPW) \citep{hogan2004instrumental}, two-way fixed effects \citep{imai2021use}, difference-in-differences (DiD) \citep{donald2007inference, callaway2021difference}, and G-estimation \citep{robins1997causal}. However, they 
are limited: IPW is sensitive to extreme weights and model misspecification \citep{waernbaum2023model}; DiD relies on parallel trends assumptions, which implies that the treatment group would have followed a similar trajectory to the control group had they not received treatment, often violated in practice \citep{roth2023s}; and G-estimation typically targets average treatment effects, offering limited insight into individual-level heterogeneity.

When it comes to causal inference and estimation of individual treatment effects, Bayesian nonparametric methods, especially Bayesian Additive Regression Trees (BART) \citep{chipman1998bayesian, chipman2010bart}, have received particular attention for their ability to flexibly estimate complex response surfaces. An important extension, Bayesian Causal Forests (BCF) \citep{hahn2020bayesian, hahn2018regularization}, improves upon standard BART by explicitly modelling prognostic and treatment effects separately, reducing regularization-induced confounding. However, both BART and BCF are inherently cross-sectional, assuming independent outcomes across individuals and time. This limits their applicability in longitudinal settings, where repeated measurements per individual introduce within-subject correlation.

To overcome these challenges, recent work has extended BART-based models to longitudinal settings. One such method is the LongBet model \citep{wang2024longbet}, which extends BCF to panel data with staggered treatment adoption by incorporating time as an input to the treatment effect function and modeling cohort-wide treatment dynamics using a Gaussian Process. While this framework is effective for capturing average treatment effect trajectories over time, it does not account for within-individual correlation and thus may miss subject-specific temporal patterns. In addition, LongBet assumes a balanced panel structure with shared observation times across individuals, and relies on imputation when follow-up schedules are irregular, which limits its applicability real-world clinical data. Separately, \cite{mcjames2024bayesian} proposed a longitudinal BCF framework (LBCF) designed to estimate treatment heterogeneity in educational settings by modelling achievement trajectories through wave-specific, cumulative growth steps. Each time interval is handled by fitting a new BART model for both prognostic and treatment effects, conditioned on the preceding wave. While this discrete-time formulation aligns well with applications where observations occur at fixed, synchronized time points, it differs fundamentally from our approach, which treats time as a continuous covariate. This distinction is crucial in clinical settings like ours, where follow-up times are irregular and individual trajectories do not align to common waves. Additionally, the need to fit separate forests at each time interval makes LBCF less scalable in scenarios involving long follow-up. Finally, \cite{yeager2022synergistic} extended BCF by incorporating a random intercept component, while \cite{yeager2019national} introduced both random intercept and slope. While these extensions were primarily designed to capture group-level structures, they could be adapted to longitudinal dynamics. However, they both do not take into account sparsity in the modelling of random effects, which is a limitation we also find in LongBet \citep{wang2024longbet} and LBCF \cite{mcjames2024bayesian}. In real-world datasets, only a subset of individuals may deviate meaningfully from the population average, and explicitly modelling this sparsity helps prevent overfitting, as well as improves the accuracy in the estimation of the longitudinal trajectory.

\subsection{Shrinkage priors}
In high-dimensional or hierarchical Bayesian models, shrinkage priors are essential for controlling overfitting, improving interpretability and prediction accuracy, and encouraging sparsity in parameter estimates \citep{chen2003random, ibrahim2011fixed}.  A wide range of shrinkage priors has been proposed, including the Bayesian lasso \citep{park2008bayesian, leng2014bayesian}, elastic net \citep{li2010bayesian}, spike-and-slab formulations \citep{ishwaran2005spike, narisetty2014bayesian},  the horseshoe prior \citep{carvalho2009handling, carvalho2010horseshoe}, and combinations of these \citep{bai2021spike}. These approaches differ in how they balance global shrinkage and local adaptivity. The Bayesian Lasso imposes a Laplace (double-exponential) prior on regression coefficients, leading to continuous shrinkage and encouraging sparsity by penalizing the absolute value of coefficients. However, it applies the same degree of shrinkage across all coefficients, which can overly penalize strong signals. The Bayesian Elastic Net combines L1 and L2 penalties, balancing sparsity and grouping effects, but suffers from similar issues of global shrinkage, lacking adaptivity across parameters. Spike-and-slab priors introduce a mixture of a point mass at zero (the "spike") and a diffuse prior (the "slab"), allowing for exact variable selection. While theoretically appealing, spike-and-slab models are computationally intensive and highly sensitive to the choice of hyperparameters. In contrast, the horseshoe prior offers a continuous global-local shrinkage formulation that applies strong shrinkage to small (likely noise) coefficients while leaving large signals largely unaffected. Its heavy tails and adaptive behavior make it particularly effective in sparse, high-dimensional settings where only a few parameters are expected to be significant. Unlike other priors, the horseshoe naturally adapts to varying signal strengths without requiring discrete selection or intensive tuning, making it well-suited for modeling random effects in complex hierarchical and longitudinal structures \citep{carvalho2009handling, carvalho2010horseshoe}. \\

\subsection{Scanner effects}
When working with neuroimaging data, there are additional factors to consider, one of the most important being scanner-related variability.
With the rise of collaborative data-sharing initiatives, such as large population neuroimaging cohorts and clinical consortia, multi-site data collection has become essential for large-scale studies due to logistical constraints and the need to capture geographic variability in subject populations \citep{laird2021large}. However, integrating data from multiple scanning sites introduces unwanted non-biological variability into imaging-derived phenotypes (IDPs), due to systematic differences between MRI scanners and acquisition protocols \citep{han2006reliability, jovicich2006reliability, goto2012influence}. This variability can obscure true biological signals, complicate statistical analysis, and limit the generalizability of findings across sites or studies. To address this, retrospective harmonization is crucial for removing non-biological variability from already collected data while preserving true biological differences in IDPs.  

The most widely used retrospective harmonization method is ComBat \citep{fortin2017harmonization}, a location-scale model that separates biological effects from scanner-specific variability. It models scanner effects as additive and multiplicative terms, estimated via empirical Bayes, and performs harmonization by centering and scaling IDPs accordingly before reintroducing the biological signal. However, despite its practical success, ComBat has notable limitations. First, it assumes scanner effects as well as biological covariates of interest influence IDPs in a strictly linear manner, which does not account for the complex, non-linear effects often observed in neuroimaging data \citep{fjell2010does}. Second, it relies on scanner or site identifiers to estimate non-biological variability, as almost all other harmonization techniques do, which can be problematic in anonymized datasets where such information is not available. 

Recent research has demonstrated that Image Quality Metrics (IQMs), which quantify intrinsic image properties such as contrast-to-noise ratio and spatial resolution, provide a richer representation of scanner effects than simple site labels \citep{esteban2017mriqc}.  Studies have shown that IQMs are strongly associated with MRI scanners and acquisition protocols, making them highly informative for harmonization \citep{esteban2017mriqc, prevot2025bartharm}.  Importantly, IQMs can be computed directly from individual MRI scans without requiring knowledge of scanner identity, making them particularly advantageous in anonymized datasets.

We have access to MS data from the NO.MS dataset, through a collaboration between Novartis, the Oxford Big Data Institute (BDI) and MS physicians \citep{dahlke2021characterisation, mallon2021advancing}. This dataset includes data from 34 Novartis MS clinical trials conducted between 2003 and January 2020, and is the largest and most comprehensive clinical trial dataset in MS with more than 35,000 patients, majority (\(>\)31,000) with relapsing–remitting MS (RRMS), with up to 15 years of follow-up data. Among these individuals, 2300 participated in placebo-controlled trials. Our goal is to precisely model continuous brain-related outcomes longitudinally, taking into account individual-level variability when present as well as account for scanner-induced biases, in order to to accurately estimate the treatment effects on brain volume loss percentages and on the change in atrophy rate over time.

Therefore, we propose a new model, the Longitudinal Bayesian Causal Forest (BCFLong), to accurately estimates longitudinal response surfaces and treatment effects. Our approach extends the Bayesian Causal Forest (BCF) framework by incorporating hierarchical random effects to model intra-individual correlation, and introduces sparsity-inducing horseshoe priors on the random effect coefficients to selectively regularize individual-level deviations by shrinking small effects aggressively while preserving large, meaningful deviations. This allows BCFLong to flexibly learn both shared population trends and individual-specific treatment trajectories.  Additionally, to address and capture scanner-induced bias, we model the prognostic BCF mean as a function of Image Quality Metrics (IQMs) extracted from the MRI scans \citep{esteban2017mriqc, prevot2025bartharm}. This is especially important in our context, as the placebo data were collected approximately 20 years prior to the treated data, leading to anticipated differences in scanner quality, imaging protocols, and acquisition settings.

By combining non-parametric fixed effects with parametric random effects, and accommodating irregular follow-ups structures, BCFLong offers a robust and general-purpose framework for individualized treatment effect estimation in complex longitudinal settings, extending beyond neuroimaging data or medical applications.

\section{Methods}
\label{sec:meth}

We consider longitudinal data collected from \( N \) individuals across multiple time points. For individual \( i = 1, ..., N \) at visit \( j = 1, ..., n_i \), where $n_i$ is the total number of observations for subject $i$, we observe a continuous outcome \( Y_{ij} \), a binary treatment assignment \( Z_{ij} \in \{0,1\} \), a covariate vector \( K_{ij} \) including the Image Quality Metrics (IQMs) for the given visit MRI scan, and a covariate vector \( W_{ij} \) containing the biological covariates which are informative for treatment.  Let \( t_{ij} \) denote the time of observation \( j \) for individual \( i \). Given the biological covariate are observed at baseline and the treatment is assigned at baseline and remains fixed over time, we drop the $j$ index for $Z$ and $W$. We extend the BCF framework by incorporating a structured random effects component, resulting in a hierarchical model, BCFLong, that can account for intra-individual correlation in repeated measurements while preserving the flexibility of the BART-based approach. We keep the fixed effects from BCF and introduce a parametric, individual-specific, random effects term to relax the independence assumption and account for the correlation structure within-individuals \citep{spanbauer2021nonparametric}. Mathematically, this can be expressed as follows:
\begin{align}\label{eq:bcflong}
    Y_{ij} &= \mu(K_{ij}, \pi_i) + \tau(W_{i}, t_{ij})Z_i + T_{ij}\alpha_i + \varepsilon_{ij} \\
    \nonumber \varepsilon_{ij} &\iidsim N(0, \sigma^2) \\
    \nonumber \mu &\sim \mathrm{SBART}_{\mu} \\
    \nonumber \tau &\sim \mathrm{SBART}_{\tau} 
\end{align}
The term $T_{ij}\alpha_i$ represents the random effects (subject-level variability) where the design vector is given by $T_{ij} = (1, t_{ij})$, and $\alpha_i = (\alpha_{i1}, \alpha_{i2})^T$ are the random effects coefficients for subject $i$, which are independent of $\varepsilon_{ij}$. $\alpha_{i1}$ and $\alpha_{i2}$ are the random intercept and slope respectively. 

In the base BCFLong model (B-BCFLong), we use a standard Gaussian prior for the slope and intercept coefficients such that $\alpha_i^B \sim N_D(0, \Sigma_{\alpha}^B)$, with an inverse-Wishart prior on variance-covariance matrix $\Sigma_{\alpha}^B$ \citep{sparapani2016nonparametric}. The inverse-Wishart (IW) distribution is parameterised by a degree of freedom parameter $\nu$ and a scale matrix $\Lambda$, specifically we set $\nu=2$ and $\Lambda=I$ to make it a weakly informative prior.  

Both $\mu(\cdot)$ and $\tau(\cdot)$\footnote{We use $\mu(\cdot)$ and $\tau(\cdot)$ as abbreviations of $\mu(K_{ij}, \pi_i)$ and $\tau(W_{i}, t_{ij})$ respectively.} are modelled with separate ensembles of soft decision trees under a Soft Bayesian Additive Regression Trees (SBART) prior \citep{linero2018bayesian}, which replaces traditional hard splits with probabilistic ones to induce smoother function estimates. The SBART prior enables flexible, nonparametric modeling of heterogeneous and potentially nonlinear effects while preserving posterior consistency and computational tractability through a sparsity-inducing regularization mechanism on the splitting rules and tree structures. We generally follow the literature to set the number of trees $m$, and the parameters for the splitting probability $\eta$ and $\beta$ parameters, which influence the probability that a node at a specific depth splits \citep{chipman2010bart,hahn2020bayesian}.  The default specification is $\eta = 0.95$ and $\beta = 2$, which already strongly favors small trees, such that the probability of a tree with 1, 2, 3, 4, and over 5 terminal nodes is 0.05, 0.55, 0.28, 0.09, and 0.03, respectively \citep{hill2020bayesian}.  We use the default 200 trees, depth penalty $\beta=2$, and $\eta=0.95$, for the prior on $\mu(\cdot)$. For the treatment effects $\tau(\cdot)$ we use only 50 trees, $\beta=3$ and $\eta=0.25$. This makes tree splitting considerably less likely, which is equivalent to shrinking more strongly toward homogeneous effects \citep{hahn2020bayesian}. We also follow BCF literature \citep{hahn2020bayesian, hahn2018regularization} and include an estimate of propensity score \( \pi_i = P(Z_i = 1 | X) \) as an input to the prognostic forest $\mu(\cdot)$ which was shown to help mitigate regularization-induced confounding and stabilize the estimation of heterogeneous treatment effects, particularly in high-dimensional or observational settings.

\subsection{Handling sparsity}\label{sec:sparsity}
To achieve sparsity in individual-level variability, we adopt the horseshoe prior, a widely used shrinkage prior in Bayesian modelling, which effectively differentiates between signal (large coefficients) and noise (small coefficients) \citep{carvalho2010horseshoe}.

We specify the prior distribution for our sparse BCFLong model (S-BCFLong) as follows:
\begin{align*}
    \alpha_i ^S &\sim \mathcal{N}_D\left(0,\   \Sigma_{\alpha}^S \right) \\
    \Sigma_{\alpha}^S &= \mathrm{diag}\left(\rho_1^2\lambda_{i1}^2,  \rho_2^2\lambda_{i2}^2\right)
\end{align*}
where \( \lambda_i = (\lambda_{i1}, \lambda_{i2}) \) are local shrinkage parameters specific to individual \( i \), \( \rho = (\rho_1, \rho_2) \) are global shrinkage parameters shared across individuals, which control the overall level of sparsity. This global-local shrinkage framework allows for aggressive shrinkage of small coefficients while preserving large coefficients, making it particularly useful for high-dimensional or sparse data. The horseshoe prior follows a hierarchical specification. For the local shrinkage parameters $\lambda_{i}$, we use
\begin{align*}
    \lambda_{id} &\sim C^+(0, 1),
\end{align*}
where \( C^+(0, a) \) denotes the standard half-Cauchy distribution with location 0 and scale $\sqrt{a}$, and $d = 1, 2$ represents the two components of the random effect (intercept and slope). This heavy-tailed prior concentrates mass near zero while allowing some coefficients to escape shrinkage, keeping strong signals intact \citep{piironen2017hyperprior}. For the global shrinkage parameter, if one has strong prior beliefs, a fixed value can be specified accordingly. Otherwise, a fully Bayesian approach can be adopted by placing a hyperprior on the global shrinkage parameter. Several choices have been proposed in the literature. \citet{carvalho2009handling} and \citet{gelman2006prior} recommended full Bayesian inference for $\rho_i$, using $\rho_{id} \sim C^+(0, 1)$. \citet{polson2010shrink} instead suggested that $\rho$ should scale with $\sigma$, such that $\rho_{id} \sim C^+(0, \sigma^2)$, in order to make inference more robust to varying number of observations and noise levels. \citet{piironen2017hyperprior, piironen2017sparsity} instead suggests that the scale $a$ should be chosen more carefully, especially for more complex scenarios where $\rho$ is only weakly identifiable from the data, and they propose constructing the scale $a$ based on the prior beliefs about the number of non-zeros in the true coefficient vector $\alpha$. This is given by $\rho_{id} \sim C^+(0, \rho_0^2)$,  where:
\begin{equation*}
    \rho_0 = \frac{N_0}{N - N_0}\frac{\sigma}{\sqrt{L}},
\end{equation*}
and $N_0$ is our prior guess for the number of relevant coefficients, $N$ is the number of individuals, i.e., total number of coefficients for each component $d$, and $L = \sum_{i=1}^Nn_i$ total number of observations across all $N$ individuals. For our model, we implement $\rho_{id} \sim C^+(0, \sigma^2)$ in the Simulation studies, following the suggestion that $\rho_{id}$ should scale with $\sigma^2$ but not using any prior knowledge about the sparsity in the data. For the real-data analysis, we conduct a sensitivity analysis by experimenting with different priors on the global shrinkage parameter.

To simplify computation and improve sampling efficiency \citep{damlen1999gibbs}, we employ the scale-mixture representation of the half-Cauchy prior \( C^+(0, a) \) proposed in \citet{wand2011mean} and \citet{makalic2015simple}, such that all conditional posterior distributions remain conjugate. For each individual \( i \), and for each component $d$ of the random effect (intercept and slope), the local and global shrinkage parameters are given hierarchical inverse-gamma priors:
\begin{align*} \label{eq:alphasparsepost}
    \lambda_{id}^2 \mid v_{id} &\sim \text{IG}\left(\frac{1}{2}, \frac{1}{v_{id}} \right) \\
    \rho_d^2 \mid \xi_d &\sim \text{IG}\left(\frac{1}{2}, \frac{1}{\xi_d} \right), \\
    v_{id}, \xi_d &\sim \text{IG}\left(\frac{1}{2}, \frac{1}{a^2} \right),
\end{align*}
where $a =1$ for $v_{id}$, and $a$ is given by either 1, $\sigma^2$, or $\rho_0^2$ for $\xi_d$.

\subsection{Causal estimands} \label{sec:estimands}
Our research questions concern two quantities of interest. The first is related to the treatment effect on the continuous outcome at a pre-specified time points. The second concerns the impact of treatment on outcome change, which captures how treatment influences the evolution of the outcome between two time points. Both estimands are formalized under the potential outcomes framework \citep{rubin1978bayesian}. We define \( Y_{ij}(1) \) and \( Y_{ij}(0) \) as the outcomes we would observe for individual \( i \) at visit \( j \), had they received treatment or control, respectively \citep{austin2011introduction, hill2011bayesian, rubin1978bayesian}. We assume the Stable Unit Treatment Value Assumption (SUTVA), which rules out interference and multiple versions of treatment \citep{imbens2015causal}. Accordingly, we only observe the potential outcome that corresponds to the realized treatment: 
\begin{equation*}
    Y_{ij} = Z_i Y_{ij}(1) + (1 - Z_i) Y_{ij}(0).
\end{equation*}
For simplicity and analogous to \citet{hahn2018regularization}, we restricted attention to mean-zero additive error representations,
\begin{align*}
    Y_{ij} &= \mu(K_{ij}, \pi_i) + \tau(W_{i}, t_{ij})z_i + T_{ij}\alpha_i + \varepsilon_{ij}, \quad \quad \quad \varepsilon_{ij} \sim N(0, \sigma^2),
\end{align*}
as shown in Equation \ref{eq:bcflong}. We then make the following assumptions, which will hold throughout the whole paper. The first is \textit{strong ignorability}, which states that conditional on observed covariates, treatment assignment is independent of the potential outcomes: $Y_{ij}(0), Y_{ij}(1) \perp Z_i | K_{ij}, W_i, t_{ij}$, for all $i=1, ..., N$ and all $j=1,...,n_i$. The second assumption is \textit{overlap}, which is necessary to estimate treatment effects everywhere in covariate space. This states that every individual has a non-zero probability of receiving treatment at baseline ($t_{ij} = 0$), i.e., $0 < Pr(Z_i=1 | W_i, t_{ij} = 0) < 1$, for all $i=1, ..., N$. Provided that these conditions hold, we can write:
\begin{equation*}
    \mathbb{E}[Y_{ij}(z_i) \mid K_{ij},W_{i},\alpha_i, t_{ij}] = \mathbb{E}[Y_{ij} \mid K_{ij},W_{i},\alpha_i, t_{ij}, Z_i = z_i],
\end{equation*}
and the treatment effect as
\begin{equation*}
    \mathbb{E}[Y_{ij} \mid K_{ij},W_{i},\alpha_i, t_{ij}, Z_i = 1] - \mathbb{E}[Y_{ij} \mid K_{ij},W_{i},\alpha_i, t_{ij}, Z_i = 0].
\end{equation*}
In our application, we are interested in the time-specific treatment effect at one or two years from baseline, that is \( t_{ij'} =1 \) and \( t_{ij} =2 \). At Year 1, this can be expressed as
\begin{align*}
     &\mathbb{E}[Y_{ij'} \mid W_{i}, K_{ij'}, \alpha_i, t_{ij'} =1,  Z_i = 1] - \mathbb{E}[Y_{ij'} \mid W_{i}, K_{ij'}, \alpha_i, t_{ij'} =1,  Z_i = 0] \\
    &= \tau(W_{i}, t_{ij'} = 1).
\end{align*}
Similarly, at Year 2, this is given by $ \tau(W_{i}, t_{ij} = 2)$. This is equivalent to the treatment effect in the BCF literature \citep{hahn2018regularization, hahn2020bayesian}, which we have followed to derive these estimands. 

Our second quantity of interest is the treatment effect on outcome change over time, defined as the difference in outcomes between two time points $t_{ij}$ and $t_{ij'}$, denoted by  $\Delta Y_{ij,j'} = Y_{ij} - Y_{ij'}$. We are still working under the potential outcome framework \citep{rubin1978bayesian} and we follow a derivation similar to \citet{mcjames2024bayesian} and \citet{angrist2009mostly}, with the difference that treatment remains constant from baseline in our application. In addition, we introduce $\overline{K}_{i} = \frac{1}{n_i}\sum_{j=1}^{n_i}K_{ij}$, which represents the average observed IQMs for subject $i$, to adjust for scanner-related variation. For each individual $i$ there are two potential differences, one observed under treatment, $\Delta Y_{ij,j'}(1)$ and one under placebo $\Delta Y_{ij,j'}(0)$. With these quantities defined, the impact of treatment on the difference in outcome during this period is given by $\Delta Y_{ij,j'}(1) - \Delta Y_{ij,j'}(0)$. However, we only ever observe one of these two, such that
\begin{equation*}
    \Delta Y_{ij,j'} = Z_i\Delta Y_{ij,j'}(1) + (1-Z_i)\Delta Y_{ij,j'}(0)
\end{equation*}

As before, we assume SUTVA, stating that $\Delta Y_{ij,j'}$ for each individual $i$, between periods $t_{ij}$ and $t_{ij'}$ is independent of any other individual treatment assignment in any period; strong ignorability, which is now written as \( \Delta Y_{ij,j'}(0), \Delta Y_{ij,j'}(1) \perp Z_i \mid W_i,\overline{K}_{i}, t_{ij}, t_{ij'} \); and overlap as before. If these conditions hold, we can write 
\begin{equation*}
    \mathbb{E}[\Delta Y_{ij,j'}(z_i) \mid W_{i}, \overline{K}_{i},\alpha_i, t_{ij,j'}] = \mathbb{E}[\Delta Y_{ij,j'} \mid W_{i}, \overline{K}_{i}, \alpha_i, t_{ij,j'}, Z_i = z_i]
\end{equation*}
We can now adapt this to our own model, shown in Equation \ref{eq:bcflong}. For the treated population ($Z_i = 1$) during an observed period $t_{ij} - t_{ij'}$, we write $\mathbb{E}[\Delta Y_{ij,j'} \mid  W_{i}, \overline{K}_{i}, \alpha_i, t_{ij'}, Z_i = 1]$ as:
\begin{align*}
     \mathbb{E} \left[  Y_{ij}  \mid W_{i}, \overline{K}_{i}, \alpha_i, t_{ij},  Z_i = 1    \right] &- \mathbb{E} \left[  Y_{ij'} \mid W_{i}, \overline{K}_{i}, \alpha_i, t_{ij'},  Z_i = 1    \right] \\
    &= \alpha_{i2}(t_{ij} - t_{ij'}) + (\tau(W_{i}, t_{ij}) - \tau(W_{i}, t_{ij'})).
\end{align*}
Similarly, for placebo population ($Z_i = 0$), we write $\mathbb{E}[\Delta Y_{ij,j'} \mid W_{i}, \overline{K}_{i} ,\alpha_i, t_{ij'}, Z_i = 0]$ as:
\begin{align*}
    \mathbb{E} \left[  Y_{ij} \mid W_{i}, \overline{K}_{i}, \alpha_i, t_{ij},  Z_i = 0    \right] &- \mathbb{E} \left[  Y_{ij'} \mid W_{i}, \overline{K}_{i}, \alpha_i, t_{ij'},  Z_i = 0    \right] \\
    &= \alpha_{i2}(t_{ij} - t_{ij'}).
\end{align*}
If we take the difference between these two we obtain our quantity of interest, that is the effect of treatment on the change in outcome between two time points $t_{ij}$ and $t_{ij'}$, 
\begin{align*}
\mathbb{E}[\Delta Y_{ij,j'} \mid  W_{i}, \overline{K}_{i}, \alpha_i, t_{ij'}, Z_i = 1] - \mathbb{E}[\Delta Y_{ij,j'} \mid W_{i}, \overline{K}_{i} ,\alpha_i, t_{ij'}, Z_i = 0] = \tau(W_{i}, t_{ij}) - \tau(W_{i}, t_{ij'})
\end{align*}
If we consider the observed period $t_{ij} - t_{ij'}$ between Year 1 and Year 2, that is \( t_{ij'} =1 \) and \( t_{ij} =2 \), we can write this quantity as the difference between the time-specific treatment effects derived earlier: $\tau(W_{i}, t_{ij}=2) - \tau(W_{i}, t_{ij'}=1)$.

\subsection{Posterior inference and posterior predictive} \label{sec:algo}

The posterior updates for the fixed effects, random effects and residual variance $\sigma^2$ are embedded into a 3-stage Gibbs sampler \citep{tan2019bayesian, spanbauer2021nonparametric}. This sampler alternately updates the scanner-related component \( \mu_{ij} \) and the residual variance \( \sigma^2 \), the biological component \( \tau_{ij} \), and the random effects coefficients $\alpha_i$.

In BART, sampling the posterior is performed through MCMC utilizing both Gibbs and Metropolis-Hastings sampling to efficiently explores the high-dimensional posterior space of tree structures and terminal node values. The updates of \( \mu_{ij} \), \( \tau_{ij} \) and \( \sigma^2 \) are adapted from traditional BART \citep{chipman1998bayesian, chipman2010bart, spanbauer2021nonparametric}. For the scanner effect \( \mu_{ij} \) update, we compute residuals \( R^\mu_{ij} = Y_{ij} - \tau(W_{i}, t_{ij})Z_i - T_{ij}\alpha_i  \), which represent the part of the outcome not explained by the biological component and the random effects. When updating \( \tau_{ij} \), we compute residuals \( R^\tau_{ij} = (Y_{ij} - \mu(K_{ij}, \pi_i) - T_{ij}\alpha_i)/Z_i \). These residuals are passed to the Bayesian back-fitting algorithm.

For the random effects coefficients $\alpha_i$ the residual is given by $R^{\alpha}_{ij} = Y_{ij} - \mu(K_{ij}, \pi_i) - \tau(W_{i}, t_{ij})Z_i$. In the base model, we follow a standard Bayesian mixed model approach to derive the posterior update equations for $\alpha_i^{B}$ \citep{spanbauer2021nonparametric}. When we instead use the sparse model, we employ the scale-mixture representation of the half-Cauchy prior and then take the posterior draws for $\alpha_i^S$ from $\mathcal{N} \left (\frac{\Psi_i^{S}}{\sigma^2} \sum_{j=1}^{n_i}T_{ij}^TR^{\alpha}_{ij}, \Psi_i^S \right )$, where $\mathcal{N}$ is a bivariate normal distribution. The precision matrix is given by
\begin{equation*}
    (\Psi_i^{S})^{-1}=\frac{1}{\sigma^2} \sum_{j=1}^{n_i}T_{ij}^TT_{ij} + (\Sigma_{\alpha}^S)^{-1} ,
\end{equation*}
where the posterior updates for $\rho$ and $\lambda_i$ for each random component $d$ are:
\begin{align*}
    \lambda_{id}^2  &\sim IG\left(1, \frac{1}{v_{id}} + \frac{\alpha_{id}^2}{2\rho_d^2} \right), \\
    v_{id} &\sim IG\left(1, \frac{1}{a_{\lambda}^2} + \frac{1}{\lambda_{id}^2} \right) \\
    \rho_d^2  &\sim IG\left(\frac{N+1}{2}, \frac{1}{\xi_d} + \frac{1}{2}\sum_{i=1}^N\frac{\alpha_{id}^2}{\lambda_{id}^2} \right), \\
    \xi_d &\sim IG\left(1, \frac{1}{a_{\rho}^2} + \frac{1}{\rho_d^2}\right).
\end{align*}
where $a_{\lambda} = 1$ and $a_{\rho} = \sigma^2$ given that we defined the priors $\lambda_{id} \sim C^+(0, 1)$ and $\rho_{d} \sim C^+(0, \sigma^2)$ on the local and global shrinkage parameters respectively.

Algorithm \ref{alg:BCFLong} outlines the iterative Gibbs sampling procedure in BCFLong. Once the for loop is completed, we obtain a full Markov chain for all updated parameters which we can use for a range of inferential tasks. We first assess the chain convergence via visual inspection of the posterior samples of the residual variance $\sigma^2$, and the random effects parameters $\rho$ and $\lambda$ to define appropriate burn-in, thinning interval and number of iterations. We then retain the posterior draws after burn-in and average across those to get a point estimate for $\hat{\tau}$, $\hat{\mu}$, and the random effect coefficents $\hat{\alpha}$, which we then use to evaluate the estimated outcome $\hat{Y}$. With the resulting sets of posterior draws we can also perform uncertainty quantification (via credible intervals), and functional summaries such as partial dependence plots, which reveal the marginal effect of covariates on either the estimated scanner effect term or the biological one. Additionally, BART also supports model-free variable selection. By tracking how frequently each covariate is used in the split rules across trees and iterations, one can estimate the relative importance of predictors. 
 
\begin{algorithm}
\caption{\textbf{BCFLong inference.} Outline of the iterative Gibbs sampling procedure used in BCFLong. BayesBackFit$^{\mu}(\cdot)$ and BayesBackFit$^{\tau}(\cdot)$ refer to the Bayesian backfitting algorithm used to update the ensembles of trees and corresponding leaf parameters for $\mu$ and $\tau$, respectively. Hyperparameters $\eta$, $\beta$, and $m$ denote the splitting probability, tree depth, and number of trees, respectively, which are specified for each ensemble.} 
\label{alg:BCFLong}
\begin{spacing}{0.9}
\begin{algorithmic}[1] 
\STATE Initialising $\tau^0_{ij} \leftarrow 0$ , and $\alpha^0_i$ $\leftarrow 0$
\STATE Initialising $\rho^{0}, \lambda_i^{0}, \xi^{0}, v_i^{0}$
\STATE
\FOR{iter $\leftarrow 1$ to max$\textunderscore$iter}
    \STATE $R^\mu_{ij} = Y_{ij} - \tau^\text{iter-1}_{ij}Z_i - T_{ij}\alpha^\text{iter-1}_i$
    \STATE Sample $\mu^\text{iter}_{ij} \text{and }(\sigma^2)^\text{iter}$ using BayesBackFit$^{\mu}$($R_{ij}^\mu, K_{ij}, (\sigma^2)^\text{iter-1}, \eta^{\mu}, \beta^{\mu}, m^{\mu}$)
    \STATE
    \STATE $R^\tau_{ij} = (Y_{ij} - \mu^\text{iter}_{ij} - T_{ij}\alpha^\text{iter-1}_i)/Z_i$
    \STATE Sample $\tau^\text{iter}_{ij}$ using BayesBackFit$^{\tau}$($ R^\tau_{ij}, W_i, t_{ij}, (\sigma^2)^\text{iter}, \eta^{\tau}, \beta^{\tau}, m^{\tau}$) 
    \STATE
    \STATE $R^{\alpha}_{ij} = Y_{ij} - \mu^\text{iter}_{ij} - \tau^{iter}_{ij}Z_i$ 
    \STATE $\alpha^\text{iter}_i \sim [\alpha_i \mid  R^\alpha_{ij}, T_{ij}, \rho^\text{iter-1}, \lambda_i^\text{iter-1}, \xi^\text{iter-1}, v_i^\text{iter-1}, (\sigma^2)^\text{iter}] $
    \STATE Sample $\rho^\text{iter}, \lambda^\text{iter}, \xi^\text{iter}, v^\text{iter}$
    \STATE
    \STATE Store posterior draws: $\mu^\text{iter}_{ij},  \tau^\text{iter}_{ij}, \alpha^\text{iter}_i, (\sigma^2)^\text{iter}, \rho^\text{iter}, \lambda_i^\text{iter}, \xi^\text{iter}, v_i^\text{iter}$
    \STATE
\ENDFOR
\end{algorithmic}
\end{spacing}
\end{algorithm}

\subsection{Simulation studies}\label{sec:sim}

In this section, we evaluate the performance of our proposed model through simulation studies designed to reflect key characteristics of the NO.MS dataset, including irregular longitudinal structure, heterogeneous treatment effects, and individual-level variability. We conduct two sets of simulations: one on fully-synthetic data to assess the accuracy of random effects estimation, and another on semi-synthetic data to evaluate the full modelling capabilities of our approach. Across all simulations, we compare the BCFLong model with sparsity-inducing priors (S-BCFLong) to the base model (B-BCFLong), which can be considered an implementation of the approaches proposed by \citet{yeager2019national} and \citet{yeager2022synergistic}, representing the closest method in the existing literature to our framework.

\subsubsection{Fully synthetic simulation}\label{sec:sim-synthetic}
The first set of experiments is designed to isolate and evaluate the model’s ability to capture individual-level variability and accurately predict outcomes in a regression setting, using a simplified BART structure. This setup focuses exclusively on assessing the benefits of incorporating random effects, with or without sparsity-inducing prior, without the added complexity of treatment effect heterogeneity. We simulate data where the true random effects coefficients, sparsity level, and outcome-generating process are known and we omit treatment terms. This setup allows us to isolate random effects performance and quantify the gains from modelling sparsity when only a subset of individuals exhibit meaningful deviation from the population trend.

We generated longitudinal data for $N=200$ individuals, each with five repeated observations, resulting in a total of $1,000$ observations.  For each observation, we generated a predictor matrix \( X \) of size $1000 \times 10$, where each covariate was sampled independently from a uniform distribution over \([0,1]\). To encode temporal structure, we replaced the 7th covariate in \( X \), denoted \( X_7 \), with a normalized observation index: values increased linearly from 0 to 1 across repeated measurements within each individual. This ensured that \( X_7 \) functioned as a proxy for time, standardized within individuals.

To incorporate individual-level heterogeneity, we generated subject-specific random intercepts \( \alpha_{i1} \sim \mathcal{N}(0,1) \) and random slopes \( \alpha_{i2} \sim \mathcal{N}(0,1) \) for each individual \( i = i, \ldots, n \).  For each observation \( j \) of individual \( i \), the random effect contribution was computed as $\gamma_{ij} = \alpha_{i1} + \alpha_{i2} \cdot X_{7, ij}$. To introduce sparsity in the random effects structure, we randomly selected a proportion (e.g., 25\%, 50\%, or 75\%) of individuals and set both their intercepts and slopes to zero. The fixed effect $\mu_{ij}$ was generated from a known non-linear function of some of the predictors, adapted from the Friedman function, to capture complex main effects. The outcome variable \( Y_{ij} \) was then computed as follows:
\[
Y_{ij} = \mu_{ij} + \gamma_{ij} + \varepsilon_{ij}, \quad \varepsilon_{ij} \sim \mathcal{N}(0, \sigma^2),
\]
where \( \sigma \) was set to 10\% of the mean of the noiseless outcome, ensuring a moderate signal-to-noise ratio. This data-generating process allows us to evaluate our model's ability to characterise sparse random effects in a simple hierarchical (mixed effects) model.

\subsubsection{Semi-synthetic simulation}\label{sec:sim-semisynthetic}
The second set of simulations is designed to more closely resemble real-world longitudinal clinical data, using observed covariates and realistic outcome structures derived from the NO.MS dataset. Here, our aim is to evaluate the model’s full capabilities, including its ability to disentangle prognostic from treatment effects, capture within-individual correlation, accurately estimate continuous outcomes, and adaptively learn sparsity in the random effects component. 

We start with the NO.MS dataset and introduce simulated outcomes while preserving the original structure of the covariates. The notation is consistent with the one introduced in the Methods section. We generated the response variable by incorporating three key components: (1) biological predictors ($W$) from NO.MS; (2) non-biological predictors ($K$) from NO.MS; and (3) individual-level random effects, capturing subject-specific deviations in both intercepts and slopes over time. The biological predictor matrix $W$ has size $2583 \times 8$ and contains biological covariates such as sex, age, Expanded Disability Status Scale (EDSS), Volume of T2 lesions, Number of gadolinium-enhancing lesions, number of relapses within one year before trial entry, duration since first symptoms, as well as follow-up time $t_{ij}$. The non-biological predictors matrix $K$ has size $2583 \times 30$ and contains the IQMs covariates used to model scanner contributions. Each individual has from 2 to 4 observations.

The coefficients for biological covariates $\beta_{\text{bio}}$ were drawn from a normal distribution with non-zero mean and higher standard deviation, \( \mathcal{N}(2, 4^2) \), to reflect both a stronger average effect and greater variability in influence on the outcome. In contrast, non-biological covariates were assigned coefficients from a distribution centered at zero with smaller standard deviation, \( \mathcal{N}(0, 0.5^2) \), simulating weak background noise with no systematic effect. The treatment indicator was encoded as \( Z_i  \in \{-0.5, 0.5\} \), to ensure symmetry and still allow the treatment effect $\tau(W_{ij})$ to be given by $W_{ij}\beta_{\text{bio}}$, for each individual \( i \) at visit $j$, where $W_{ij}$ includes the follow-up time $t_{ij}$.

To introduce individual heterogeneity, for each individual \( i \), we simulated random intercepts \( \alpha_{i2} \sim \mathcal{N}(0.5, 3^2) \) and slopes \( \alpha_{i1} \sim \mathcal{N}(0.5, 2^2) \), defining the random effect contribution as \( \gamma_{ij} = \alpha_{i1} \cdot t_{ij} + \alpha_{i2} \), where \( t_{ij} \) is the follow-up time. To induce sparsity, we randomly selected a proportion of individuals (e.g., 25\%, 50\%, 75\%) and set both their intercepts and slopes to zero.  We then computed the observed outcome as:
\[
Y_{ij} = \gamma_{ij} + (W_{ij}\beta_{\text{bio}})Z_i + K_{ij}\beta_{\text{nonbio}} + \varepsilon_{ij}, , \quad \varepsilon_{ij} \sim \mathcal{N}(0, \sigma^2),
\]
where \( \sigma \) was set to 10\% of the mean of the noiseless outcome.

This simulation structure provides a controlled yet realistic environment for assessing the full modelling capabilities of our approach.

\subsection{Performance evaluation}
For all simulation studies, we focus on three key aspects: the accuracy of parameter recovery for both fixed and random effects, the quality of inference, and the model’s posterior predictive performance on held-out observations. To support this evaluation, we adopt a subject-level partitioning strategy of the data. Specifically, we randomly sample $x$ individuals and assign one of their observations to an \textit{held-out set}. The remaining observations as well as all observations of the remaining individuals are used for model fitting to obtain posterior samples. This ensures that we have new, unseen data to evaluate posterior predictive performance but also that each individual's random effect $\alpha_i$ can be estimated during inference.\\

Starting from the the fully-synthetic dataset described in Section \ref{sec:sim-synthetic}, we are interested in isolating and evaluating estimation of random effects to see whether using sparsity inducing prior is beneficial. We model the fixed effects component using a single SBART ensemble over all 10 covariates in \( X \) described in Section \ref{sec:sim-synthetic}, specified with 100 trees, and the random effects (slope and intercept) are modelled either with a base Gaussian prior (B) or a sparsity-inducing prior (S). Importantly, even if we still refer to this model as BCFLong, we have omitted the BCF-style fixed mean decomposition and only use one tree ensemble plus the random effects component. This is equivalent to \citet{spanbauer2021nonparametric} mixed BART model, with the addition of the sparsity inducing prior. We evaluate model performance along two dimensions: parameter estimation and posterior predictive. For parameter estimation, we assess the recovery of individual-specific random intercepts ($\alpha_1$) and slopes ($\alpha_2$), total subject-specific random effect contribution $\gamma$, and outcome $Y$ by computing Root Mean Squared Error (RMSE) between posterior means and the true generative parameters. For both $\gamma$ and $Y$, we additionally assess the quality of the posterior predictive performance by evaluating RMSE, 95\% credible interval coverage and interval width on the held-out dataset.  Performance is assessed over 1,000 independent realizations of the simulated dataset. For each run, we use 5,000 Gibbs sampling iterations, discarding the first 1,000 as burn-in. Convergence was confirmed through visual inspection of posterior trace plots for the residual variance $\sigma^2$, conducted on a random subset of realizations.

For the semi-synthetic simulation set-up described in Section \ref{sec:sim-semisynthetic} we are interested in evaluating full modelling capabilities, including not only outcome and random effects, but also treatment effects and prognostic (scanner) mean. The model we are using is the BCFLong presented in Equation \ref{eq:bcflong}, using one SBART ensemble of 200 trees $\mu(K_{ij}, \pi_i)$ to model the IQMs covariates $K_{ij}$; a separate SBART ensemble of 50 trees $\tau(W_{ij})$ to model the time-specific treatment effect on the biological covariates $W_{ij}$ including visit time $t_{ij}$; and the random effects (slope and intercept) are modelled under either a base Gaussian prior (B) or a sparsity-inducing prior (S). We also implement a vanilla BCF, keeping $\mu(K_{ij}, \pi_i)$ and $\tau(W_{ij})$. 

We again evaluate model performance along two dimensions: parameter estimation and posterior predictive accuracy. For parameter estimation, we compute RMSE between the true, simulated, value and the posterior means to assess the recovery of the outcome $Y$, the treatment effect $\tau$, the prognostic (scanner) mean $\mu$, and the total subject-specific random effect contribution $\gamma$. For $Y$, $\tau$ and $\gamma$, we additionally assess the quality of the posterior predictive performance by evaluating RMSE on the held-out dataset, as well 95\% credible interval coverage and interval width for $Y$. Whenever we evaluate RMSE for the treatment effect $\tau$ we can also refer to this as the Precision in Estimating Heterogeneous Effects (PEHE): $PEHE = \sqrt{\frac{1}{L}\sum_{i=1}^N\sum_{j=1}^{n_i}(\tau_{ij} - \hat{\tau}_{ij})^2}$. Here, we do not restrict our attention to treatment effects at pre-defined time point, such as Year 1 or 2, but evaluate the ability of BCFLong to recover the time-specific effects at the visit time $t_{ij}$ that is present in the dataset. Performance is assessed over 1,000 independent realizations of the simulated dataset. For each run, we use 10,000 Gibbs sampling iterations, discarding the first 3,000 as burn-in. Convergence was confirmed through visual inspection of posterior trace plots for the residual variance $\sigma^2$, conducted on a random subset realizations.

\begin{figure}[h!]
\begin{center}
\includegraphics[width = 6in]{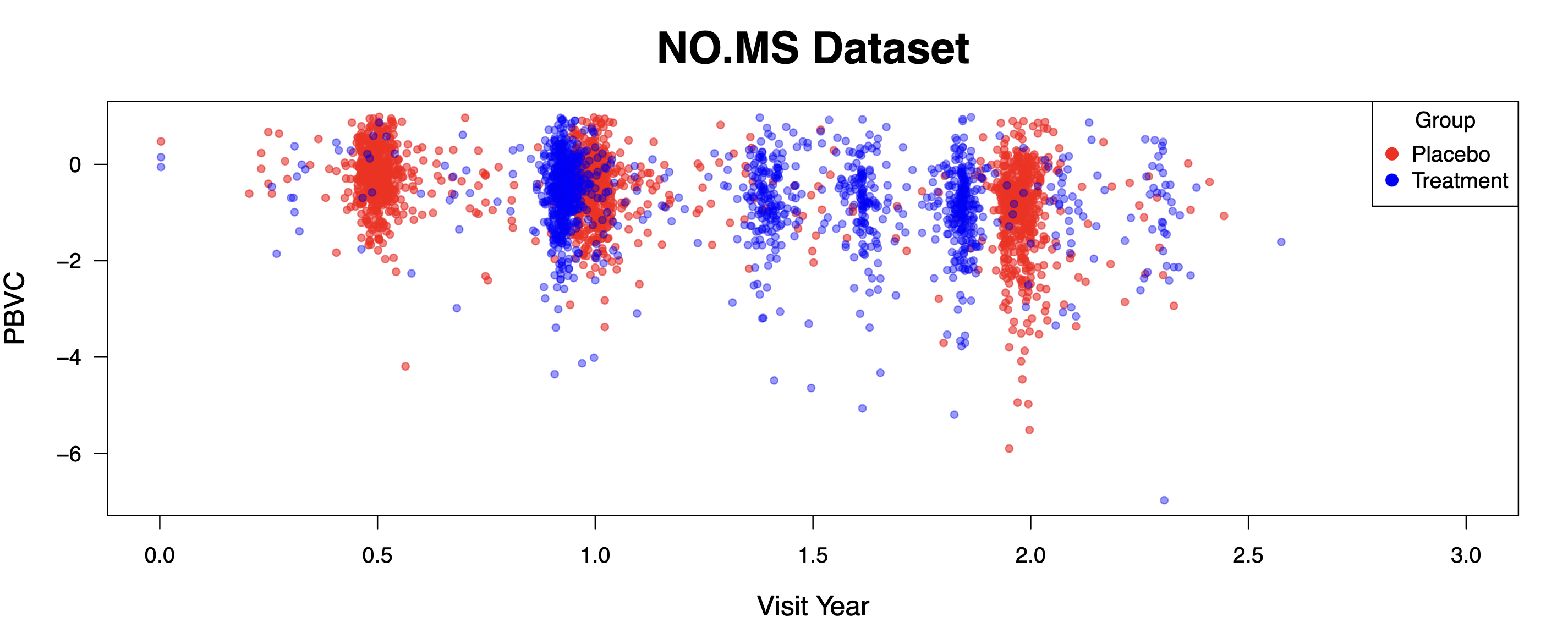}
\end{center}
\caption{\textbf{Patient-level PBVC trajectories in the NO.MS dataset.} 
Scatter plot of percentage brain volume change (PBVC) over visit years, stratified by treatment group. Each point represents a patient visit, with blue denoting patients on treatment and red denoting those on placebo. The plot highlights the irregular timing of visits and variability in PBVC trajectories across groups.}
\label{fig:NOMSvisits}
\end{figure}

Finally, for real-data analysis we evaluated BCFLong on the NO.MS dataset to estimate the Percentage Brain Volume Change (PBVC) continuous outcome at Year 1 and 2 vs baseline. We retained 3151 observations from 1406 unique individuals in NO.MS, including 648 on placebo and the rest receiving treatment. Each individual has from 2 to 4 observations. From Figure \ref{fig:NOMSvisits} we can observe high level of irregularity in the visits patterns, which is what motivated us to develop a method that can model response surface accurately despite this irregular follow-up times.

We model the prognostic mean $\mu(\cdot)$ on the IQMs and the propensity score using 200 trees, and the treatment effect $\tau(\cdot)$ on the biological covariates and the follow up time, measured as Years since baseline, using 50 trees. The biological covariates are namely sex, age, MS type, Normalised Brain Volume, Expanded Disability Status Scale (EDSS), Volume of T2 lesions, Number of gadolinium-enhancing lesions, number of relapses within one year before trial entry, and duration since first symptoms. All these biological covariates are measured at baseline. The IQMs are extracted using MRIQC \citep{esteban2017mriqc} from the T1w MRI data, providing 56 features per anatomical image. These fall into four categories: (1) noise characterization, (2) spatial information distribution, (3) artifact detection (e.g., INU artifact, motion-related signal leakage), and (4) general image properties (e.g., tissue distribution, volume overlap, sharpness). Each observation will have its own set of IQMs values extracted from the corresponding MRI. 

We run the Gibbs sampler for 100,000 iterations, with thinning of 10 and discard the first 3000 for burn-in post-thinning. The treatment indicator was encoded as \( Z_i  \in \{-0.5, 0.5\} \), for placebo and treated respectively, to ensure symmetry \citep{hahn2018regularization}. We perform posterior inference on the full NO.MS dataset and report results of both models, the base BCFLong model (B), which can be seen as an implementation of \cite{yeager2019national, yeager2022synergistic}, adapted to our own problem setting, and the sparse BCFLong model (S), which is our main contribution. For the latter, we conduct a sensitivity analysis by experimenting with different priors on the global shrinkage parameter, as described in Section~\ref{sec:sparsity}. Additionally, for comparison, we implemented a vanilla BCF model \citep{hahn2018regularization, hahn2020bayesian}, which does not include random effects components.
 
\section{Results}

\subsection{Simulations}

Tables~\ref{tab:synthetic1} and~\ref{tab:synthetic2}  summarize the performance of BCFLong on the fully-synthetic simulation set-up describe din Section \ref{sec:sim-synthetic} with either base (B) Gaussian prior or sparsity-inducing horseshoe prior (S) on the random effects, across varying levels of sparsity. Overall, both models achieve accurate recovery of outcome and individual-level effects and well-calibrated predictions, but the use of a sparsity-inducing prior yields clear improvements. In terms of parameter estimation, S-BCFLong consistently achieves lower RMSE, especially for random effects (\( \alpha_1, \alpha_2, \gamma  \)) as sparsity increases. For posterior predictive performance, S-BCFLong shows clear improvements in both \( \gamma \) and $Y$, particularly under higher sparsity as expected. At 50\% sparsity, the RMSE for predicting \( \gamma \) decreases from 0.29 (B) to 0.21 (S). In the most extreme case (75\% sparsity), the credible interval width for \( \gamma \) is reduced by over 50\% (from 0.99 to 0.46). Crucially, these gains extend to the outcome $Y$, where S-BCFLong not only improves predictive accuracy, but also yields narrower and more calibrated predictive intervals. Taken together, strong performance on both parameter estimation and posterior predictive inference indicates that the sparse model is not only better at recovering meaningful structure from the data, but also generalises better to new observations.

\begin{table}[h]
    \centering
    \renewcommand{\arraystretch}{1.1}
    \begin{tabular}{l l cc|cc}
        \hline
        & & \multicolumn{2}{c|}{No sparsity} & \multicolumn{2}{c}{25\% sparsity}  \\
        \cline{1-6}
        & & B & S &  B & S  \\
        \hline
        Parameter Estimates & $Y$ & 0.08 & 0.09   & 0.09 & 0.09 \\
        (RMSE) & $\alpha_1$  & 0.35 & 0.32 & 0.27 & 0.26 \\
        & $\alpha_2$  & 0.47 & 0.43 & 0.41 & 0.39 \\
        & $\gamma$  & 0.15 & 0.16 & 0.14 &  0.12 \\
        \hline
        Posterior Predictive & $Y$ RMSE  & 0.31 & 0.32 & 0.30  &  0.29 \\
        & $Y$ 95\% Coverage   & 0.89 & 0.89  & 0.86 & 0.88 \\
        & $Y$ 95\% CI Width & 1.02 & 1.08 & 1.01 & 0.92 \\
        \cline{2-6}
        & $\gamma$ RMSE  & 0.30 & 0.28 & 0.26 & 0.25 \\
        & $\gamma$ 95\% Coverage  & 0.92 & 0.95 & 0.94 & 0.96 \\
        & $\gamma$ 95\% CI Width & 1.09 & 1.02 & 1.01 & 0.87 \\
        \hline
    \end{tabular}
    \caption{\textbf{Performance comparison in fully-synthetic simulations with low sparsity.} Comparison of BCFLong with either base (B) Gaussian prior or sparsity-inducing (S) horseshoe prior on random effects, under two settings: no sparsity and 25\% sparsity.}
    \label{tab:synthetic1}
\end{table}

\begin{table}[h]
    \centering
    \renewcommand{\arraystretch}{1.1}
    \begin{tabular}{l l cc|cc}
        \hline
        & &  \multicolumn{2}{c|}{50\% sparsity} & \multicolumn{2}{c}{75\% sparsity}  \\
        \cline{1-6}
        & & B & S &  B & S  \\
        \hline
        Parameter Estimates & $Y$ & 0.10 & 0.08  & 0.11 & 0.09 \\
        (RMSE) & $\alpha_1$  & 0.36 & 0.33 & 0.35 & 0.26 \\
        & $\alpha_2$ & 0.27 & 0.25 & 0.25 & 0.19 \\
        & $\gamma$  & 0.13 & 0.11 & 0.13 &  0.08 \\
        \hline
        Posterior Predictive & $\gamma$ RMSE & 0.29 & 0.21 & 0.21 & 0.16 \\
        & $\gamma$ 95\% Coverage & 0.93 & 0.96 & 0.93 & 0.98 \\
        & $\gamma$ 95\% CI Width & 0.99 & 0.98 & 0.99 & 0.46 \\
        \cline{2-6}
        & $Y$ RMSE & 0.28 & 0.25 & 0.26 & 0.22 \\
        & $Y$ 95\% Coverage & 0.84 & 0.89 & 0.90 & 0.86 \\
        & $Y$ 95\% CI Width & 0.99 & 0.73 & 0.99 & 0.53 \\
        
        \hline
    \end{tabular}
    \caption{\textbf{Performance comparison in fully-synthetic simulations with high sparsity.} Comparison of BCFLong with either base (B) Gaussian prior or sparsity-inducing (S) horseshoe prior on random effects, under two settings: 50\% sparsity and 75\% sparsity. }
    \label{tab:synthetic2}
\end{table}

Importantly, even in the no-sparsity setting, S-BCFLong maintains competitive performance across all metrics, slightly improving predictive accuracy and interval calibration relative to B-BCFLong. This demonstrates the ability of the sparsity inducing prior in S-BCFLong to adaptively shrink non-informative coefficients toward zero, while preserving signal in non-zero cases. \\

We then moved to the semi-synthetic scenario described in Section \ref{sec:sim-semisynthetic} with real-world NO.MS covariates but known ground truth. The results in Tables~\ref{tab:semisynthetic_low} and~\ref{tab:semisynthetic_high} clearly illustrate the limitations of the vanilla BCF model and the benefits of incorporating subject-level variability through BCFLong. Across all sparsity levels, both BCFLong variants, base (B) and sparse (S), substantially outperform vanilla BCF in both parameter estimation and posterior predictive performance. Notably, modelling individual-level random effects leads to major improvements in estimating not only the outcome $Y$, but also the treatment effect $\tau$. For the latter, this is particularly striking: BCF struggles to recover heterogeneous effects with precision even when just a few individuals have non-zero random effects, while BCFLong, by explicitly accounting for intra-individual correlation, achieves lower PEHE. For example, under 75\% sparsity, PEHE drops from 0.73 with BCF to 0.52 with B-BCFLong and 0.16 with S-BCFLong. This demonstrates that failing to account for within-subject structure limits both inference and prediction, even for quantities not directly tied to the random effects.

\begin{table}[h]
    \centering
    \renewcommand{\arraystretch}{1.1}
    \begin{tabular}{ll ccc|ccc}
        \hline
        & & \multicolumn{3}{c|}{No sparsity} & \multicolumn{3}{c}{25\% sparsity}  \\
        \cline{1-8}
        & & BCF & B & S & BCF & B & S  \\
        \hline
        Parameter Estimates & $\tau$   & 1.19 & 0.86 & 1.05 &  1.06 & 0.76 & 0.60 \\
        (RMSE/PEHE) & $\mu$ & 1.01 & 0.77 & 0.83   & 0.84 & 0.57 & 0.34 \\
        & $\gamma$   & - & 0.97 & 1.01 & - & 0.80 & 0.62\\
        & $Y$   & 2.90 & 0.48 & 0.49 & 2.52 & 0.49 & 0.43 \\
        \hline
        Posterior Predictive & PEHE   & 1.19 & 0.88 & 1.11 & 1.02 & 0.75 & 0.60 \\
        & $\gamma$ RMSE   & -  & 1.22 & 1.29 & - & 1.02 & 0.89 \\
        & $Y$ RMSE & 3.10 & 0.93 & 0.95 & 2.63  & 0.89 & 0.84 \\
        & $Y$ 95\% Coverage & 0.37 & 0.85 & 0.87 & 0.48 & 0.86 & 0.89  \\
        & $Y$ 95\% CI Width   & 2.89 & 2.52 & 2.67 & 2.72 &  2.42 & 2.27 \\
        \hline
    \end{tabular}
    \caption{\textbf{Performance comparison on semi-synthetic data with low sparsity.} Evaluation of vanilla BCF, base BCFLong (B), and sparse BCFLong (S) models under no sparsity and 25\% sparsity conditions. Modelling individual-level variability (BCFLong) proves beneficial under low sparsity, as expected. Incorporating sparsity-inducing priors (S) consistently yields better performance, even when no sparsity is present.}
    \label{tab:semisynthetic_low}
\end{table}

\begin{table}[h]
    \centering
    \renewcommand{\arraystretch}{1.1}
    \begin{tabular}{ll ccc|ccc}
        \hline
        & & \multicolumn{3}{c|}{50\% sparsity} & \multicolumn{3}{c}{75\% sparsity}  \\
        \cline{1-8}
        & & BCF & B & S & BCF & B & S  \\
        \hline
        Parameter Estimates & $\tau$   & 0.96 & 0.66 & 0.28 & 0.77 & 0.52 & 0.14\\
        (RMSE/PEHE)& $\mu$   & 0.65 & 0.43 & 0.17 & 0.48 & 0.30 & 0.09\\
        & $\gamma$   & - & 0.64 & 0.44 & - & 0.49 & 0.29\\
        & $Y$   & 2.07 & 0.49 & 0.37 & 1.44 & 0.41 & 0.30\\
        \hline
        Posterior Predictive & PEHE  & 0.91 & 0.67 & 0.28 & 0.73  & 0.52 &  0.16\\
        & $\gamma$ RMSE   & -  & 0.81 & 0.68 & - & 0.60 & 0.45 \\
        & $Y$ RMSE & 2.13 & 0.70 & 0.69 &  1.51 & 0.59 & 0.50 \\
        & $\hat{Y}$ 95\% Coverage & 0.59 & 0.90 & 0.91  & 0.73 & 0.92 & 0.94 \\
        & $\hat{Y}$ 95\% CI Width  & 2.24 & 2.30 & 1.95 & 1.81 &  2.10  & 1.28 \\
        \hline
    \end{tabular}
    \caption{\textbf{Performance comparison on semi-synthetic data with high sparsity.} Evaluation of vanilla BCF, base BCFLong (B), and sparse BCFLong (S) models under 50\% and 75\% sparsity conditions. As sparsity increases, modelling individual-level variability (BCFLong) remains advantageous, but the benefits of incorporating sparsity-inducing priors become even more pronounced.}
    \label{tab:semisynthetic_high}
\end{table}

Focusing now on the comparison between B-BCFLong and S-BCFLong, we observe that the inclusion of a sparsity-inducing prior substantially enhances parameter estimation, particularly under moderate and high sparsity. Across all levels of sparsity, S-BCFLong yields lower RMSE for all key parameters. The advantages of S-BCFLong extend clearly to posterior predictive performance. Across all settings, it achieves lower RMSE, narrower and more calibrated credible intervals, on the held-out outcome $Y$. At 75\% sparsity, for example, $Y$ RMSE drops from 0.59 (B) to 0.50 (S), and the 95\% credible interval width shrinks from 2.10 to 1.28. Despite this narrowing, coverage improves slightly from 0.92 to 0.94, suggesting that the sparse model captures uncertainty more efficiently without sacrificing calibration. Improvements are also observed for the posterior prediction of $\gamma$ and $\tau$, where RMSE consistently decreases under S-BCFLong. These results indicate that incorporating sparsity leads to more reliable and precise predictive distributions, which is critical for downstream use of the model.

A particularly noteworthy finding is the strong performance of S-BCFLong in estimating subject-specific random effects.  Across all sparsity regimes, S-BCFLong outperforms B-BCFLong in terms of parameter estimate RMSE, suggesting that the sparsity prior helps suppress spurious individual variability. This highlights the regularizing benefit of sparsity  while still preserving signal for subjects with meaningful deviations. In fact, even under the no-sparsity regime, where all individuals exhibit random effects, the sparse model performs competitively with the base model. Although one might expect the sparse prior to underperform when sparsity is absent, these results highlight its adaptive behaviour: it does not eliminate significant effects, but rather shrinks only where supported by the data. 

\begin{figure}[h!]
\begin{center}
\includegraphics[width = 6in]{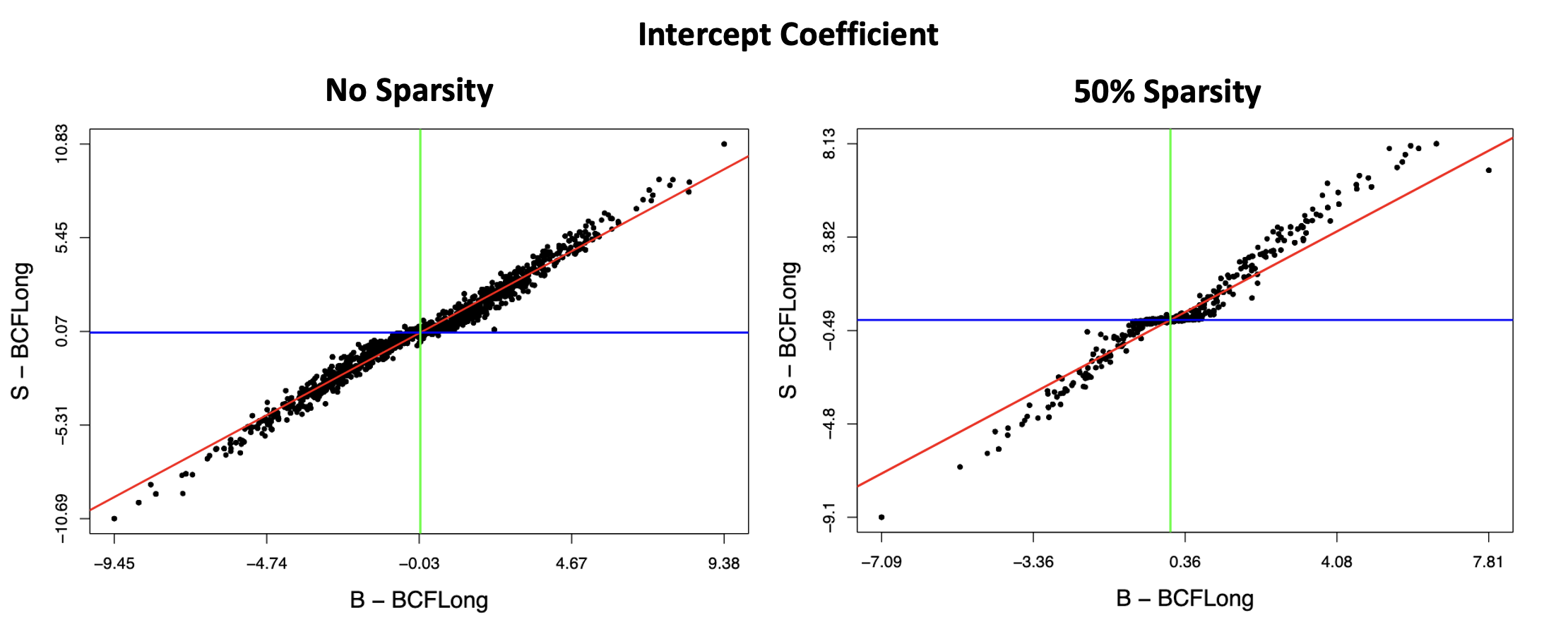}
\end{center}
\caption{\textbf{Comparison of subject-level intercept estimates between B-BCFLong and S-BCFLong.} 
Each point represents a subject’s estimated random intercept under the base (x-axis) and sparse (y-axis) BCFLong models. The red line denotes perfect agreement. The left panel shows results under no sparsity; the right panel under 50\% sparsity. The sparsity prior in S-BCFLong strongly shrinks small or spurious coefficients toward zero while allowing larger, signal-bearing coefficients to "escape" shrinkage.}
\label{fig:intercept_semi}
\end{figure}

Figure~\ref{fig:intercept_semi} provides a visual comparison of the parameter estimates for random effects intercepts between B-BCFLong and S-BCFLong under two scenarios: no sparsity (left panel) and 50\% sparsity (right panel). Each point represents a subject-level intercept estimate. The red diagonal line indicates perfect agreement between the models, while the green and blue lines mark zero on the $x$- and $y$-axes, respectively. Under both settings, the estimates are closely aligned, but the deviation from the identity line increases slightly with sparsity, reflecting the shrinkage effect introduced by the sparsity prior. This is a characteristic feature of the horseshoe prior, which strongly shrinks small or spurious coefficients toward zero while allowing larger, signal-bearing coefficients to "escape" shrinkage. As sparsity increases, this selective regularization becomes more pronounced, suppressing noise while preserving interpretable individual-level heterogeneity.

Taken together, these results demonstrate a clear pattern: vanilla BCF is inadequate for longitudinal settings. Modelling individual-level variability via BCFLong yields substantial improvements in both parameter estimation and predictive performance of the outcome response as well as the treatment effects. As the degree of sparsity in the random effects increases, the sparse BCFLong model (S) clearly outperforms its base counterpart (B), producing more accurate posterior mean estimates and more efficiently calibrated predictive distributions. Crucially, S-BCFLong maintains competitive performance in the no-sparsity regime, where all individuals have non-zero random effects. Despite the lack of true sparsity, the model adaptively regularizes only where appropriate. This robustness across varying levels of complexity is particularly important for applications like our MS case study, where precise modelling of time-varying and individual-specific treatment responses is essential.

\subsection{Applications to NO.MS} \label{sec:nrealdata}
We now apply BCFLong to the NO.MS dataset to estimate the Percentage Brain Volume Change (PBVC) continuous outcome at Year 1 and 2 vs baseline. We will first discuss the model predictive performance across random effects, scanner effects, and overall response modelling. Then we will move to the causal analysis.

When using the sparse version of BCFLong and experimenting with the different priors on the global shrinkage parameter, as described in Section~\ref{sec:sparsity}, we found no substantial differences between the priors $\rho_{id} \sim C^+(0, 1)$ and $\rho_{id} \sim C^+(0, \sigma^2)$, as $\sigma^2$ is always estimated to be around $0.7$. In contrast, specifying $\rho_{id} \sim C^+(0, \rho_0^2)$ resulted in noticeably stronger shrinkage. However, to fully understand the implications of varying prior beliefs about the number of non-zero entries in the true coefficient vector, further experiments are needed. Therefore, we report all main results using the prior $\rho_{id} \sim C^+(0, \sigma^2)$, which provides a balanced level of regularization.

\begin{figure}
\begin{center}
\includegraphics[width = 6in]{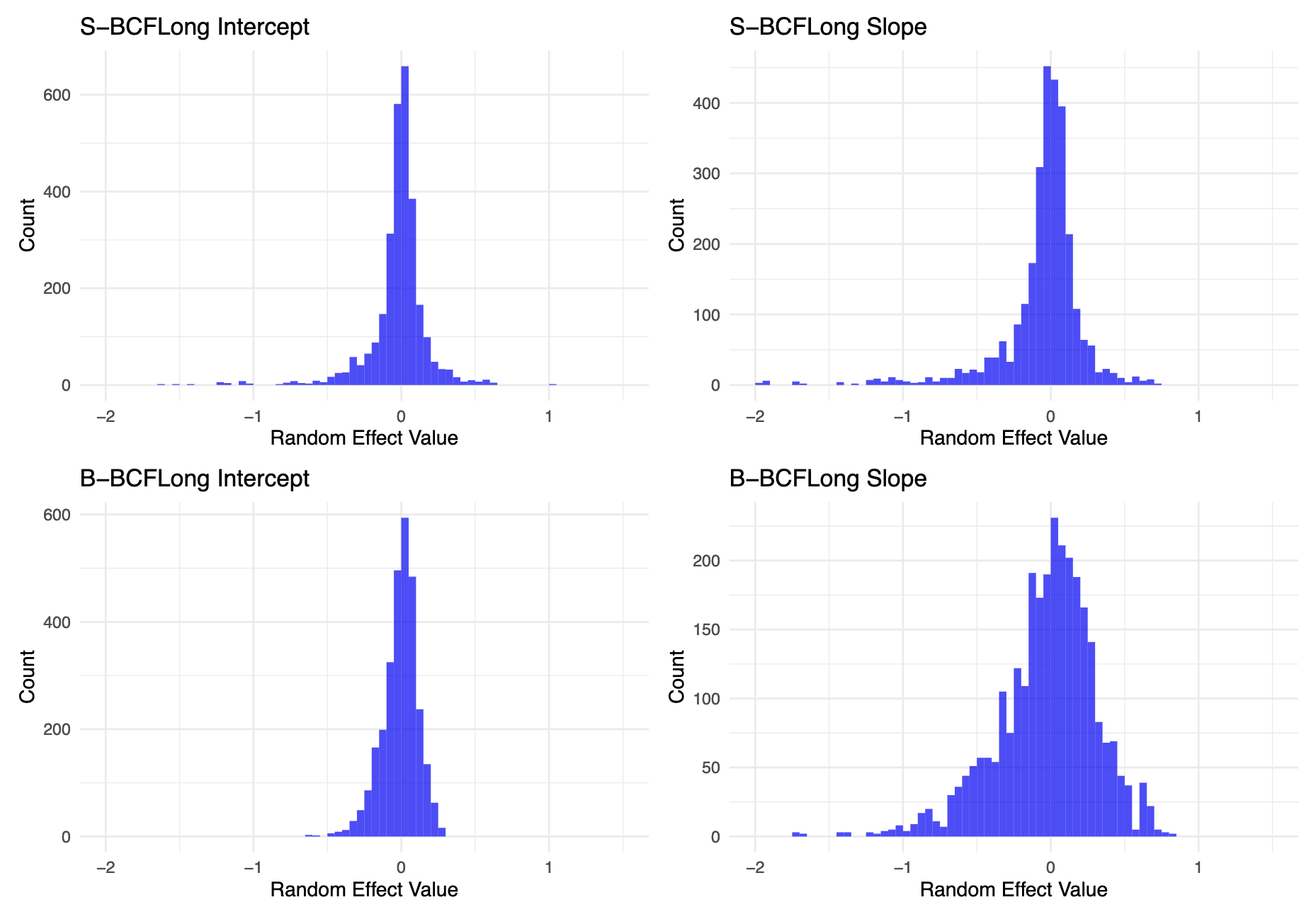}
\end{center}
\caption{\textbf{Comparison of estimated individual-level random effects under two prior settings in the BCFLong model.} The top row shows histograms under the sparsity-inducing horseshoe prior (S-BCFLong), while the bottom row uses a flat baseline prior (B-BCFLong). Left panels display intercept effects ($\alpha_1$), and right panels display slope effects ($\alpha_2$). The horseshoe prior yields more concentrated estimates near zero, especially for the slope, reflecting stronger regularization.}
\label{fig:hist_alphas}
\end{figure}

Figure \ref{fig:hist_alphas} shows the distribution of estimated individual-level random effects under the two BCFLong variants. The top row corresponds to the sparse model (S-BCFLong) using a horseshoe prior, while the bottom row shows results under the flat prior (B-BCFLong). In both cases, we separately examine the random intercepts ($\alpha_1$) and slopes ($\alpha_2$). The horseshoe prior induces stronger shrinkage on the slope term, yielding a tighter distribution around zero. This reflects the prior’s adaptivity in suppressing noise while preserving large, informative effects. The flat prior allows for broader dispersion as expected.

\begin{figure}
\begin{center}
\includegraphics[width = 5in]{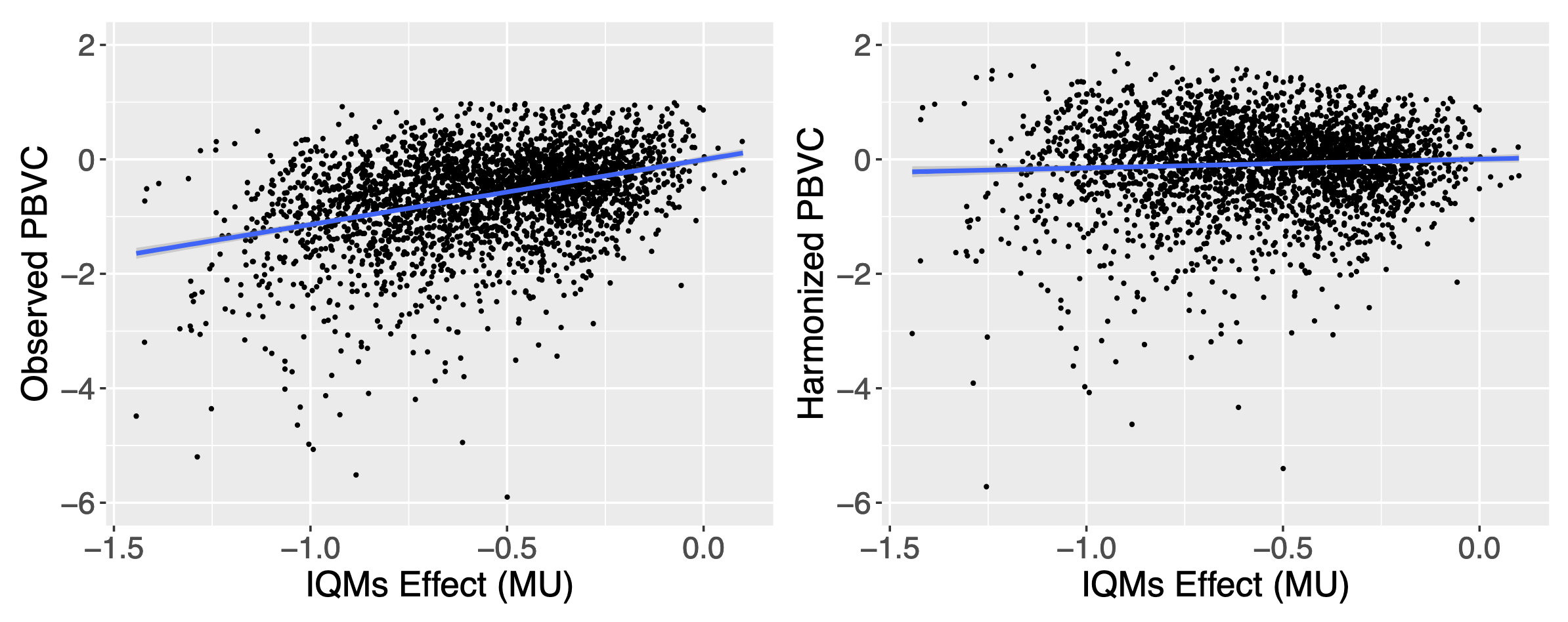}
\end{center}
\caption{\textbf{Scanner-related harmonization.} 
Scatter plots showing the relationship between the estimated scanner effects and PBVC, before (left) and after (right) harmonization. In the observed data, scanner-related variability is clearly associated with PBVC values. After harmonization using the approach proposed in BARTharm \citep{prevot2025bartharm}, this association is substantially reduced.}
\label{fig:pbvc_harm}
\end{figure}

For the scanner effects term, Figure \ref{fig:pbvc_harm} (left) shows that scanner-related variability from IQMS is clearly associated with PBVC values. By modelling $\mu(\cdot)$ as a function of IQMs, we are able to isolate scanner-related variability in PBVC to characterise more accurately underlying biological variation, as demonstrated in \cite{prevot2025bartharm}. This is demonstrated by Figure \ref{fig:pbvc_harm} (right), where removing the estimated scanner effect from the observed outcome yields a near-flat relationship.

\begin{figure}
\begin{center}
\includegraphics[width = 6.5in]{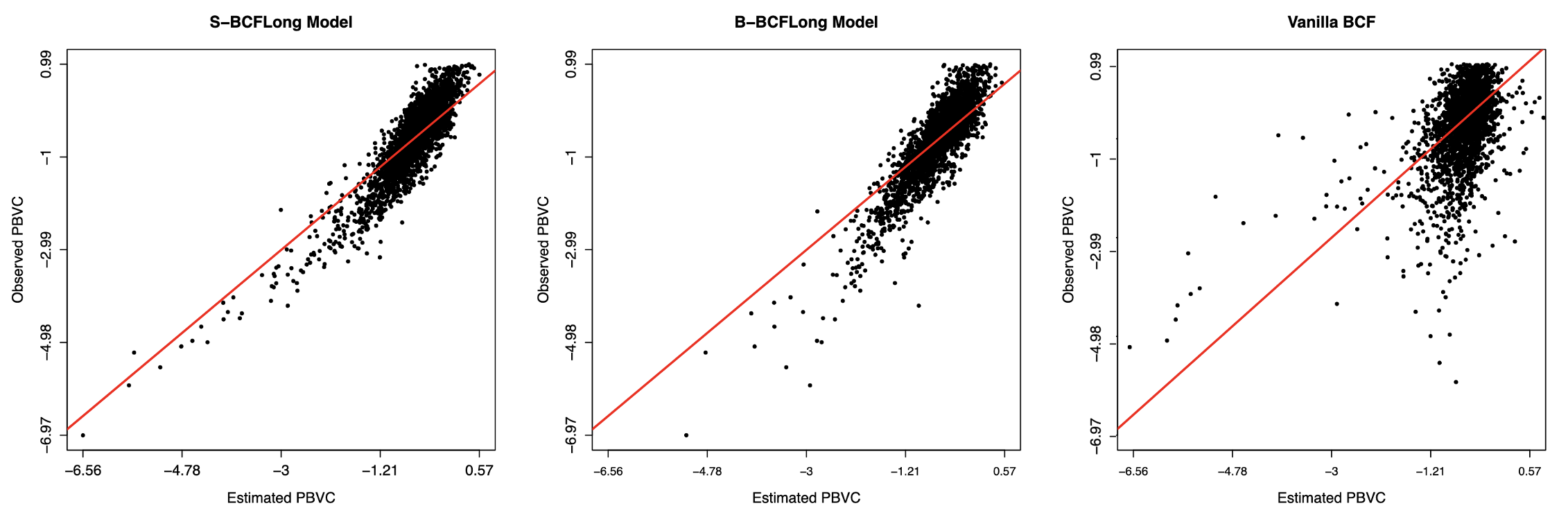}
\end{center}
\caption{\textbf{Model fit comparison on the NO.MS dataset.} 
Visualization of predicted versus observed percentage brain volume change (PBVC) over time across three models: sparse BCFLong (S-BCFLong), base BCFLong (B-BCFLong), and vanilla BCF. BCFLong demonstrates an improved fit to the data compared to vanilla BCF, and the sparse variant shows a slightly better fit than the base version (RMSE 0.41 vs 0.46).}\label{fig:NOMSfit}
\end{figure}

Moving now to the overall predictive performance, Figure~\ref{fig:NOMSfit} compares model fit to the observed data using posterior mean estimates from the vanilla BCF approach and our proposed BCFLong framework, in both its base (B-BCFLong) and sparse (S-BCFLong) variants. The figure highlights an improvement in fit achieved by BCFLong, which explicitly models subject-specific intercepts and longitudinal dependencies. While the difference between the base and sparse versions is visually subtle, RMSE values indicate a slight improvement for the sparse model: B-BCFLong achieved an RMSE of 0.46, while S-BCFLong was 0.41; vanilla BCF RMSE was instead 0.61. \\

\begin{figure}
\begin{center}
\includegraphics[width = 6.5in]{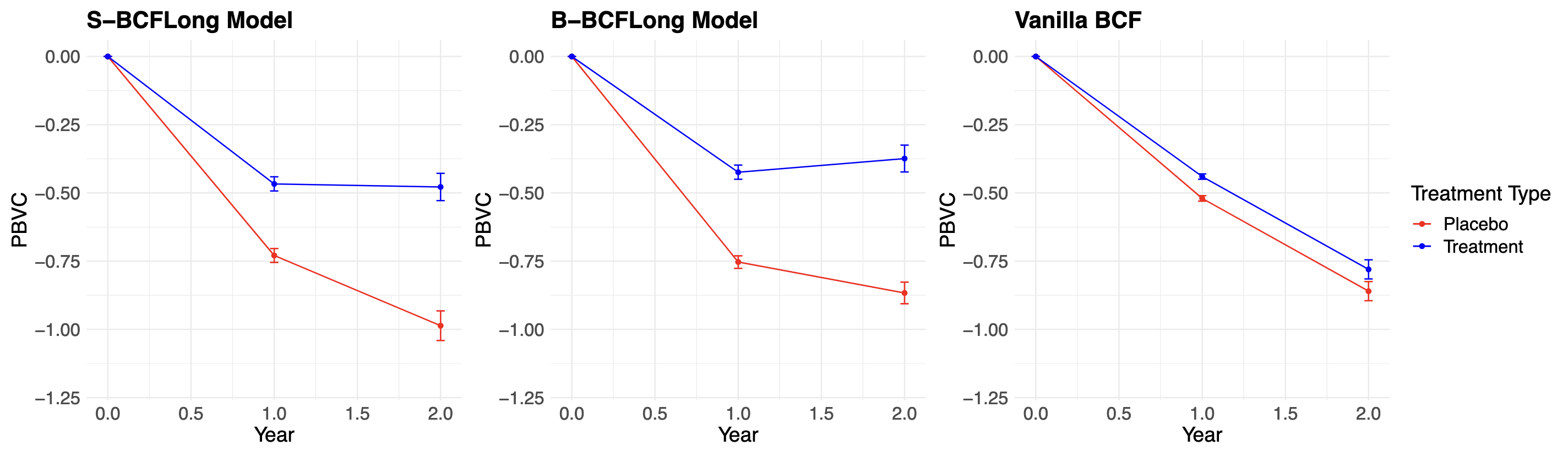}
\end{center}
\caption{\textbf{Estimated PBVC trajectories by model and treatment group.} 
Estimated percentage brain volume change (PBVC) over two years for placebo (red) and treatment (blue) arms, shown across three models: S-BCFLong (left), B-BCFLong (center), and vanilla BCF (right). }
 \label{fig:NOMSpbvc}
\end{figure}

Our causal quantities of interest are the time-specific treatment effect on PBVC at Year 1 and 2 and the treatment effect on PBVC change between Year 1 and 2, as defined in Section~\ref{sec:estimands}. Figure~\ref{fig:NOMSpbvc} shows the estimated PBVC trajectories over two years, comparing BCFLong (both base and sparse) with the vanilla BCF model. Each point represents the estimated PBVC averaged across all individuals, based on counterfactual inference, with 95\% credible intervals reflecting posterior uncertainty. All models support a significant treatment effect, with the treated group showing consistently lower PBVC in magnitude (i.e., reduced atrophy) than the placebo group at both time-points. However, the magnitude of the estimated average treatment effect varies notably across models. 

 \begin{table}[h]
    \centering
    \renewcommand{\arraystretch}{1.1}
    \begin{tabular}{l ccc}
        \hline
        CATE & \multicolumn{3}{c}{Estimates} \\
        & S-BCFLong & B-BCFLong & Vanilla BCF \\
        \hline
        Year 1:  $\tau(W_{i}, t_{ij'} = 1)$ & 0.26 [0.09, 0.44] &  0.33 [0.15, 0.52] & 0.11 [0.03, 0.22]\\
        Year 2:  $\tau(W_{i}, t_{ij} = 2)$ & 0.81 [0.62, 1.01] & 0.96 [0.74, 1.18] & 0.13 [0.05, 0.19] \\
        Long:  $\tau(W_{i}, t_{ij} = 2) - \tau(W_{i}, t_{ij'} = 1)$  & 0.55 [0.41, 0.69] & 0.63 [0.48, 0.78] & 0.02 [0.01, 0.04] \\
        \hline
    \end{tabular}
    \caption{\textbf{Estimated Conditional Average Treatment Effects (CATE) at Year 1, Year 2, and the longitudinal difference between these time points}, based on three models: Sparse BCFLong (S-BCFLong), Base BCFLong (B-BCFLong), and Vanilla BCF. Values represent posterior means with 95\% credible intervals. The cross-sectional, vanilla, BCF fails to identify a longitudinal effect, highlighting the importance of modelling within-subject correlations to avoid underestimating time-varying treatment effects. }
    \label{tab:causal_estimates}
\end{table}

Table \ref{tab:causal_estimates} shows the Conditional Average Treatment Effects (CATE) estimated by the different models, with 95\% credible interval (CI). BCFLong models estimate substantially larger treatment effects than the vanilla BCF, with the base version (B-BCFLong) consistently producing slightly higher estimates than the sparse variant (S-BCFLong). Additionally, BCFLong estimates a longitudinal treatment effect on PBVC change between Year 1 and Year 2, which is also evident in the flattening trend between these two time-points (Figure~\ref{fig:NOMSpbvc}).  According to S-BCFLong, on average, treatment is expected to reduce the increase in individuals' PBVC (atrophy) between Year 1 and 2 by between 0.41 and 0.69 units, and similarly for B-BCFLong by between 0.48 and 0.78. On the other hand, BCF fails to capture this longitudinal effect. This suggests that by ignoring within-subject correlations and longitudinal dependencies, the cross-sectional BCF model may oversimplify the true trajectory of PBVC change, potentially underestimating time-varying treatment effects.

If we focus our comparison between S-BCFLong with B-BCFLong, we observe similar PBVC trends over time but, the sparse model yields more plausible biological estimates. Specifically, S-BCFLong avoids seemingly implausible increases in brain volume and maintains a nearly linear decline for the placebo group, consistent with expectations. This distinction is meaningful, as both models estimate near-identical scanner effects, so differences in outcomes reflect differences in treatment and random effect estimates.

\begin{figure}
\begin{center}
\includegraphics[width = 5.5in]{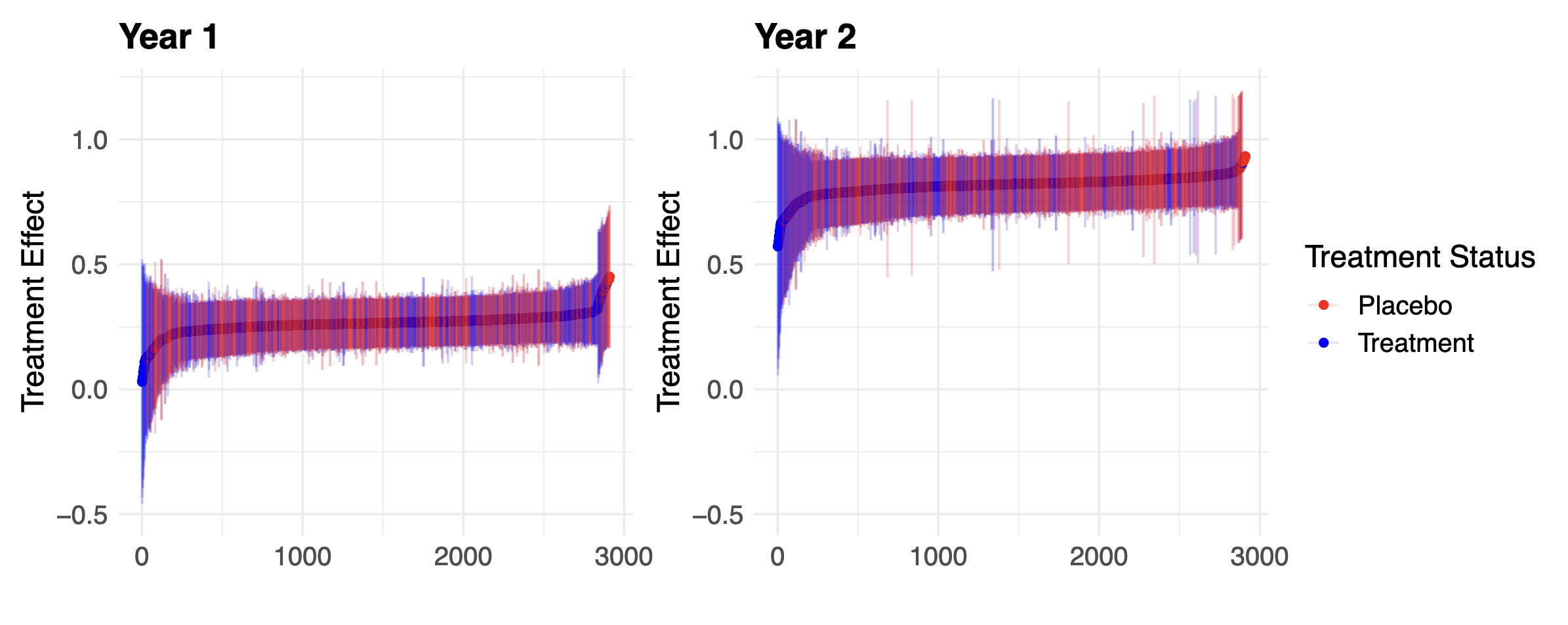}
\end{center}
\caption{\textbf{Estimated individual treatment effects.} 
Posterior mean treatment effects and 95\% credible intervals for each individual at Year 1 (left) and Year 2 (right), estimated using S-BCFLong. Estimates are sorted and coloured by actual treatment status. Treatment effects increase over time and remain positive across nearly all individuals, suggesting consistent benefit with varying magnitude. We obtained a very similar plot for B-BCFLong.}
\label{fig:tau_comparison}
\end{figure}

We next focus on the individual conditional treatment effect (ICATE) at Year 1 and 2, which is shown in Figure~\ref{fig:tau_comparison} for the sparse BCFLong model. Subjects are grouped by the true treatment assignment and we provide 95\% Credible Intervals. Across both time points, nearly all individuals exhibit a positive treatment effect, with confidence bounds remaining above zero. Moreover, the magnitude of the effect increases from Year 1 to Year 2, consistent with the pattern observed in PBVC trajectories (Figure~\ref{fig:NOMSpbvc}). These findings suggest not only an overall positive effect of treatment but also meaningful individual-level variation in responsiveness as some patients benefit more than others. 

\section{Discussion}
\label{sec:conc}

Our work introduced BCFLong, a hierarchical extension of Bayesian Causal Forests to model longitudinal data with individual-level variability. A central feature of our approach is the incorporation of sparsity-inducing priors on random effects. This enhances model flexibility, enabling it to dynamically adapt to the presence or absence of individual deviations and temporal slopes, without requiring strong prior assumptions. Unlike traditional mixed-effects models that impose fixed parametric structures, BCFLong learns from the data whether such individual-level terms are necessary. This is particularly valuable in longitudinal studies, where the extent and structure of within-individual variability are often unknown. Additionally, our model is capable of isolating non-biological variability, which might be affecting the outcome of interest but is not related to the treatment effect response, to characterise more accurately underlying biological variation and avoid biasing estimation. This is highly relevant for neuroimaging studies to remove scanner effects. The result is a robust data-driven modelling framework that can accommodate a wide range of complex, real-world longitudinal patterns.

Our simulation studies provide strong support for this approach. In fully synthetic scenarios, both the base and sparse versions of BCFLong accurately recovered random effects, achieving low RMSE and credible interval coverage close to the nominal level. However, the sparse model consistently outperformed the base model as the degree of sparsity increased, in both posterior estimates and posterior predictive. In the more realistic semi-synthetic simulations using covariates from the NO.MS dataset, BCFLong significantly improved outcome prediction and treatment effect estimation compared to vanilla BCF with no random effects. Moreover, incorporating a sparsity inducing prior on these random effects showed even superior performance, suggesting that adaptive shrinkage leads to better generalization in the presence of subject-specific heterogeneity. Subsequently, the application to the NO.MS dataset further demonstrated the benefits of modelling individual-level variability with BCFLong. The model achieved improved fit over vanilla BCF, particularly in its ability to capture longitudinal patterns in Percentage Brain Volume Change (PBVC). Additionally, by modelling random effects with a data-adaptive sparsity prior, we obtained more accurate and realistic outcome estimates and treatment effect trajectories through counterfactual estimation. 

While our discussion has primarily focused on PBVC to illustrate the model's structure and causal inference capabilities, we have also applied BCFLong to a broader set of imaging-derived phenotypes (IDPs) relevant to Multiple Sclerosis, including regional gray matter volumes, to help assess whether the treatment effect is diffuse or localized and investigate the biological covariate driving this effect and variability. Across these applications, BCFLong enabled the identification of biologically meaningful and clinically consistent effects, highlighting its utility beyond a single outcome and reinforcing its potential as a general-purpose framework for neuroimaging-based causal modelling in MS.

However, further experiments are needed to better understand the shrinkage behavior of the sparse model, particularly in cases where we noticed extreme regularization from the global shrinkage prior. While this may reflect genuine structure in the data, it could also be a symptom of overshrinking, potentially suppressing meaningful individual variability, and thus requires further investigation. 

Despite these open questions, our results provide strong support for the effectiveness of the sparse BCFLong as a general and flexible framework for causal inference. The method achieves accurate estimation of treatment effects, robust outcome prediction, and interpretable modelling of individual variability, which are features critical to real-world applications such as clinical trials and longitudinal cohort studies. The ability to incorporate random effects selectively when the data supports them and the robustness to settings with irregular follow-up offer a key advantage over existing methods. Moreover, the general flexibility of the model has sparked interest within our research group, where it is now being adopted to extend other methodological frameworks, including normative modelling and longitudinal harmonization.

\section*{Data Availability}

For NO.MS data, the reader is able to request the raw data (anonymized) and related documents (e.g., protocol, reporting and analysis plan, clinical study report) of all the studies that underlie the modelling results reported in this article by connecting to \\
https://www.clinicalstudydatarequest.com and signing a Data Sharing Agreement with Novartis. These will be made available to qualified external researchers, with requests reviewed and approved by an independent review panel on the basis of scientific merit.

\section*{Funding}

E.P. is a doctoral student at the University of Oxford, supported by the Oxford EPSRC Centre for Doctoral Training in Health Data Science (EP/S02428X/1).

\section*{Declaration of Competing Interests}

EP, TEN, CCH, and HG declare no competing interest. DAH is an employee and shareholders of Novartis Pharma AG.

\newpage

\bibliographystyle{agsm}

\bibliography{Bibliography-MM-MC}

\end{document}